# PCSI — The Platform for Content-Structure Inference

*Caleb Malchik*
*Joan Feigenbaum*

Yale University
New Haven, Connecticut 06511
{caleb.malchik,joan.feigenbaum}@yale.edu

*ABSTRACT*

The Platform for Content-Structure Inference (PCSI, pronounced "pixie") facilitates the sharing of information about the process of converting Web resources into structured content objects that conform to a predefined format. PCSI records encode methods for deriving structured content from classes of URLs, and report the results of applying particular methods to particular URLs. The methods are scripts written in Hex, a variant of Awk with facilities for traversing the HTML DOM.

## 1. Introduction

The dominant web and mobile applications of today are characterized by top-down control over not just remote services but the interfaces necessary to access them. Users have scant influence over these systems, especially when acting in isolation, and high switching costs render alternatives nonviable.

PCSI, the *Platform for Content-Structure Inference*, aims to address these inadequacies by enabling user-controlled client software to present information that is typically accessed through service-controlled web applications. PCSI does not depend on the participation of websites; rather, it adapts to changing page formats through the collection and sharing of scripts that specify how to extract various data, imposing a structure that is not present in the raw HTML.

The interoperability enabled by PCSI is crucial to allow users to experience the benefits of new applications without sacrificing access to content distributed through established channels. Doctorow [1] defines *adversarial interoperability* as "interoperability that is undertaken against the wishes of the originators of some product or service"; he also includes *cooperative* and *indifferent* interoperability in his taxonomy. The approach taken by PCSI may be characterized as *client-driven interoperability*, because the system's relation to websites need not be adversarial, but interoperability is achieved without their support and is driven entirely by client software.

Eventually, adoption of PCSI and related technologies may induce content and service providers to accommodate new formats and protocols directly, reducing the need for interoperability measures and freeing up resources for the benefit of users as well as services. In this way, PCSI aims to build a bridge from the current morass of web applications to a new ecosystem of applications that are rationally designed and controlled by users.

We now present the core components of PCSI, including *structured content objects* that expose the structure of content in a machine-readable form; *Hex*, the language PCSI uses to convert web pages into structured content objects; and the various types of PCSI *records* that are shared among PCSI instances to help select appropriate scripts and content objects. News articles are used as an example application throughout.

## 2. Structured Content Objects

The PCSI system is designed to expose the structure of content in a machine-readable format, such that client software offering an interface to the content may provide the greatest benefit to the user. At the highest level, PCSI organizes content into *structured content objects*, each corresponding to a URL and having a type such as *article*:

```
(article
  (headline "BBC complains to Apple over misleading shooting headline")
  (date "1734112342")
  (author "Graham Fraser")
  (body ...))
```

PCSI objects and records are S-expressions following the format defined by Rivest [2]. An S-expression is a byte string or a parenthesized list of S-expressions: The article object above is a list consisting of one string ("article") followed by four lists. Some strings are quoted in our presentation for clarity, but the quoting encodes no semantics and is not present in the canonical form used for storage, transmission, and hashing. In the canonical form, whitespace between strings is removed and each string is prefixed by a decimal length and a colon:

```
(7:article(8:headline56:BBC complains to Apple over misleading shooting
headline)(4:date10:1734112342)(6:author13:Graham Fraser)(4:body...))
```

The given object includes fields for the *headline*, publication *date* (in seconds since January 1, 1970), *author*, and *body* of the article. The *body* consists of a list of *paragraphs*, *subheadings*, and *images*:

```
(body
  (image (url "https://ichef.bbci.co.uk/news/480/cpsprodpb/4e44/live/5688\
  a220-b972-11ef-a2ca-e99d0c9a24e3.jpg.webp") (caption "Luigi Mangione \
  is accused of murdering healthcare insurance CEO Brian Thompson"))
  (paragraph "The BBC has complained to Apple after the tech giant's new \
  iPhone feature generated a false headline about a high-profile murder \
  in the United States.")
  (paragraph "Apple Intelligence, launched in the UK earlier this week"
  (link "https://www.apple.com/uk/newsroom/2024/12/apple-intelligence-is-\
  available-today/") ", uses artificial intelligence (AI) to summarise \
  and group together notifications.")
  ...
  (subheading "'Embarrassing' mistake")
  (paragraph "Apple says one of the reasons people might like its \
  AI-powered notification summaries is to help reduce the interruptions \
  caused by ongoing notifications, and to allow the user to prioritise \
  more important notices.")
  ...)
```

A *paragraph* is a list of strings and *links*; A *link* has only a URL and no link text, and may be rendered with a clickable button or a numerical reference. A *subheading* simply has the text of the subheading. An *image* has a *url* and an optional *caption*.

The format of content objects is defined for a given type by agreement among parties in a (hypothetical) PCSI ecosystem. Each object type with an agreed-upon format provides a basis for PCSI applications that implement an interface to content of that type. Our chosen format for news articles targets simplicity, both for expository purposes and to minimize the cost of developing and maintaining reliable interface software. (The article objects presented here are used in PCSINews, a news reader application to be described in a future publication.)

## 3. Processing HTML with Hex

The central function of PCSI is to convert data (usually HTML) downloaded from a URL into a structured content object of a given type. This conversion process is dictated by small programs written in a language called Hex by users of varying technical ability and shared throughout the ecosystem. A PCSI application collects scripts from various sources and automatically selects scripts that are likely to work for a given URL.

The *article* object above is the result of downloading an HTML document from the URL `https://www.bbc.com/news/articles/cd0elzk24dno` and filtering it through a script that is given in full in Appendix A. Here is the top-level routine:

```
BEGIN {
    walk(root())
    n = split("headline date author body", fields)
    for (i = 1; i <= n; i++) {
        f = fields[i]
        if (!A[f])
            error("missing " f)
        x = x object(f, A[f])
    }
    print object("article", x)
}
```

Hex is Awk [3] with additional built-in functions for traversing an HTML DOM (Document Object Model) tree. In the code above, `root` is a new built-in function that returns the ID of the root node of the tree. The functions `walk`, `error`, and `object` are defined by the user elsewhere in the script.

The code above traverses the DOM tree, storing field values in the associative array `A`. It then checks that each field has been successfully assigned and builds a list of objects in the string `x` that becomes the contents of an *article* object. Here is the implementation of `object`:

```
function object(s, x) {
    if (!x) return ""
    return "(" verbatim(s) x ")"
}
function verbatim(s) { return length(s) ":" s }
```

The `object` function expects `x` to be a concatenated list of canonical S-expressions. It converts the object name `s` into length-prefixed form and returns a canonical S-expression for the object.

The processing of the HTML document to assign values to fields happens in the `walk` function and makes heavy use of the built-in functions that Hex adds to Awk, described in the table below.

| Function | Value Returned |
|---|---|
| `root()` | the ID of the root node (typically 1) |
| `parent(n)` | the ID of the parent of node $n$ |
| `sister(n)` | the ID of the next sibling of node $n$ |
| `children(n)` | the ID of the first child of node $n$ |
| `type(n)` | the type of node $n$, expressed as a string: one of `ELEMENT`, `TEXT`, `COMMENT`, `DECLARATION`, `PROCINS`, or `ROOT` |
| `name(n)` | the name of the HTML element at node $n$ |
| `text(n)` | the text contents of node $n$, where $n$ is a text node |
| `attr(n,s)` | the value of the $s$ attribute in the HTML element at node $n$ |
| `selmatch(n,s)` | 1 if the CSS selector $s$ matches node $n$, 0 otherwise |
| `seconds(s)` | the number of seconds since 00:00:00 GMT, Jan. 1, 1970 corresponding to $s$ interpreted as a date string |

The functions that operate on tree nodes return 0 if no valid return value is available; for example, if the given node ID is not valid or if `children` is called on a leaf node. `Seconds` returns the empty string if it fails to interpret its argument as a date, as 0 represents midnight on January 1, 1970.

- 4 -At the start of execution, Hex parses its input as an HTML document into a tree of nodes, each assigned a unique numerical ID. The input is not otherwise available to the program, so the "pattern-action" facilities of Awk are not used. Our implementation borrows code from Bert Bos's XML-HTML-utils*, which is also the origin of many of the names of the new built-in functions.

The `walk` function below illustrates a typical in-order traversal of an HTML document.

```
function walk(n,    json, s) {
    if (!n) return
    if (children(n)) {
        if (name(n) == "h1") {
            only("headline", text(children(n)))
        } else if (name(n) == "script" &&
            attr(n, "type") == "application/ld+json") {
            json = text(children(n))
            s = between(json, "\"datePublished\":\"", "\"")
            only("date", seconds(s))
            s = between(json, "\"author\":[", "]")
            s = between(s, "\"name\":\"", "\"")
            only("author", s)
        } else if (selmatch(n, "main#main-content article div p")) {
            A["body"] = A["body"] paragraph(n)
        } else if (selmatch(n, "main#main-content article div h2")) {
            A["body"] = A["body"] subheading(n)
        } else if (selmatch(n, "main#main-content article figure")) {
            A["body"] = A["body"] image(children(n))
        }
        walk(children(n))
    }
    walk(sister(n))
}
```

First note the extra function parameters `json` and `s`. These are local variables, distinguished from function arguments by the presence of several blanks. This is the normal way to define local variables in Awk and we cannot be blamed for it.

The string of `if-else` statements checks the current node for a number of conditions, and if one is satisfied then the fragment of HTML descendent from the current node is relevant to some field in the final object: If the current node is an `h1` element, then a child text node is expected to contain the headline of the article. The user-defined `only` function assigns a string as a canonical S-expression to the global array `A`, producing an error if it has already been set.

The *date* and *author* are found in a JSON-LD object within a `script` element. We extract them using a simple `between` function that returns a string found within its first argument between its second and third arguments. In our example, the `datePublished` string is `2024-12-13T17:52:22.647Z`, and `seconds` interprets this as a date by trying a range of possible date formats until one matches.

The rest of the code is concerned with the relatively complex task of assembling the `body` out of chunks that match certain CSS selectors. See Appendix A for the full implementation.

## 4. PCSI Records

We have described how PCSI can use a script to produce a structured content object from a URL, but how does PCSI know which script to run, and what if a script fails to produce an accurate structured content object? PCSI records are produced and shared by users and organizations to help PCSI applications select appropriate scripts and content objects to provide a user-experience that is as seamless as possible.

---

__________
* https://www.w3.org/Tools/HTML-XML-utils/

### 4.1. Rule Records

*Rule* records suggest a script to run for a class of URLs matching a pattern. They are produced manually by technical users or staff, usually in conjunction with the writing of a Hex script. Here is a typical *rule* record:

```
(rule
  (source |WRlcbFQcgwfx2i0edo1vIoDJhN8hetX0xkw1QrBBEaQ=|)
  (timestamp "1735608269")
  (pattern 57:https?://(www\.)?bbc\.com/news/articles/[0-9a-z][0-9a-z]*)
  (script-hash |CY7Iwrrw5i7MyjV7Zqdwf2Tj0Hb3iCsJF4Sv6jcrUyw=|)
  (object-type article))
```

The *pattern* is a regular expression using the same syntax found in Awk; we represent it here in length-prefixed "verbatim" mode for clarity.

The *source* is an identifier for who produced the record; here it is a fingerprint of a public key. The *timestamp* tells when the record was produced.

The *script-hash* is a cryptographic hash of the script being suggested. A *script* field my be included in addition to or instead of the *script-hash*, containing the text of the script itself. The *object-type* is the type of object produced by the script.

### 4.2. Inference Records

*Inference* records describe the running of a script for a given URL and its result: in abstract terms, an attempt by PCSI to infer the structure of the content. They are produced automatically when PCSI runs a script, and may be automatically published or shared with a list of contacts. *Inference* records share many of the fields found in *rule* records:

```
(inference
  (source |JMFb3PAURhU2D2FUWT3XO1TV0Vm5Lf6rL63ctK0G10s=|)
  (timestamp "1735610039")
  (url "https://www.bbc.com/news/articles/cd0elzk24dno")
  (script-hash |CY7Iwrrw5i7MyjV7Zqdwf2Tj0Hb3iCsJF4Sv6jcrUyw=|)
  (object-type article)
  (object-hash |11wG9jfDHtChbXEvFaTHd+eDxzyE1FxVPazbJUNy6zs=|)
```

An *inference* record such as this one denotes that data was downloaded from the given *url* at the given *timestamp*, then run through the script identified by the *script-hash*, yielding an object identified by the *object-hash*. If the script yields an error, an *error* field containing the error message is used and *object-type* and *object-hash* are omitted.

The fields *script* and *object* may also be used instead of an identifying hash, but care must be taken when sharing a record containing an *object* field to avoid infringing the copyright of the work represented in the object. For this reason, *object* fields are not included in records that are shared automatically by PCSI software.

### 4.3. Perception Records

As an *inference* record is typically generated automatically, it represents no claim that the object produced is appropriate to the given URL. *Rule* records provide some such suggestion, but they are imprecise because they apply to an entire class of URLs. *Perception* records allow users to express when a specific object does or does not match the content at a URL:

```
(perception
  (source |hTVgZNA4cqxL7Rebi76DGKtnqWJr5V0NciiO4U4WUmU=|)
  (timestamp "1735610400")
  (url "https://www.bbc.com/news/articles/cd0elzk24dno")
  (object-type article)
  (object-hash |11wG9jfDHtChbXEvFaTHd+eDxzyE1FxVPazbJUNy6zs=|)
  (valid "1"))
```

The only field we have not yet encountered is *valid*, denoting with a 1 or 0 whether the object accurately represents the information relevant to the *object-type* present at the *url*. The ``1'' means that the source has checked the object against the URL as rendered in a browser, or however she thinks the publisher intended the content to be consumed, and believes that the object encodes the article-relevant information presented at the BBC.com address.

This somewhat tedious task may be forced upon users if a website updates its format and the scripts available to the PCSI software no longer work. Capturing these moments of confusion and sharing the resulting *perception* records can help the broader PCSI ecosystem adapt and fix problems as they arise. In practice, a script such as the one presented here will generate an error if a page is not formatted as expected, producing an *inference* record with an *error* field that can be shared automatically, and no errant content object will be presented to the user.

## 5. The PCSI Ecosystem and Beyond

The framework presented here provides a foundation for the development of applications that are maximally accountable to the user. To build a PCSI application, one must establish a format for structured content objects for a certain type of content, and identify means to obtain URLs that serve the desired content: Our PCSINews application, for example, gets lists of URLs serving article content from user-supplied RSS feeds.

With sufficient availability of accurate PCSI records, applications can overcome network effects that would normally force users to rely on a modern browser that hands control of the interface over to the publisher of the content. Ultimate control over a user interface rests with the developer of the software, so PCSI attempts to minimize the software complexity necessary to render content or to participate in a PCSI network. Then even nontechnical users can exercise control over their computing by choosing software that behaves as they wish, or enlisting a technically skilled associate to produce the same.

PCSI is best suited to deployment among a group of users, some of whom have the technical skills and motivation to write Hex scripts and *rule* records. Organizations such as the *data co-ops* described in [4] would be ideally suited to supporting PCSI software, but PCSI is lightweight and flexible enough to offer value in a wide range of deployment settings.

The main limitation of PCSI is that it does not support use cases that involve uploading data to a service. Applications such as chat, discussion forums, and microblogging could be supported by a more general framework that offered a way to describe the interaction with a server; in PCSI, this interaction is reduced to fetching content from a URL.

**Appendix A: A Hex script for BBC articles**

```
BEGIN {
    walk(root())
    n = split("headline date author body", fields)
    for (i = 1; i <= n; i++) {
        f = fields[i]
        if (!A[f])
            error("missing " f)
        x = x object(f, A[f])
    }
    print object("article", x)
}
function object(s, x) {
    if (!x) return ""
    return "(" verbatim(s) x ")"
}
function verbatim(s) { return length(s) ":" s }
function walk(n,    json, s) {
    if (!n) return
    if (children(n)) {
        if (name(n) == "h1") {
            only("headline", text(children(n)))
        } else if (name(n) == "script" &&
            attr(n, "type") == "application/ld+json") {
            json = text(children(n))
            s = between(json, "\"datePublished\":\"", "\"")
            only("date", seconds(s))
            s = between(json, "\"author\":[", "]")
            s = between(s, "\"name\":\"", "\"")
            only("author", s)
        } else if (selmatch(n, "main#main-content article div p")) {
            A["body"] = A["body"] paragraph(n)
        } else if (selmatch(n, "main#main-content article div h2")) {
            A["body"] = A["body"] subheading(n)
        } else if (selmatch(n, "main#main-content article figure")) {
            A["body"] = A["body"] image(children(n))
        }
        walk(children(n))
    }
    walk(sister(n))
}
function only(k, v) {
    if (!v) return
    if (A[k])
        error("multiple values for " k ":\nprevious: " A[k] "\ncurrent: " v)
    A[k] = verbatim(v)
}
function between(s, before, after,    start) {
    start = index(s, before) + length(before)
    s = substr(s, start, length(s) - start)
    return substr(s, 1, index(s, after) - 1)
}
function paragraph(n) {
    return object("paragraph", hypertext(children(n)))
}
function subheading(n) {
    return object("subheading", hypertext(children(n)))
}
```

```
function hypertext(n,    a) {
    i = hypertext1(n, a, 1)
    if (a[1])
        sub(/^[ \t]+$/, "", a[1])
    if (a[i])
        sub(/[ \t]+$/, "", a[i])
    return csexp(a)
}
# i: where the last item was added
function hypertext1(n, a, i) {
    if (!n) return i
    if (type(n) == "TEXT") {
        if (a[i,1]) # last part was a link
            i++
        a[i] = a[i] text(n)
    } else {
        i = hypertext1(children(n), a, i)
        if (name(n) == "a" && attr(n, "href")) {
            i++
            a[i,1] = "link"
            a[i,2] = attr(n, "href")
        }
    }
    i = hypertext1(sister(n), a, i)
    return i
}
function csexp(a, i,    j, x) {
    if (i)
        i = i SUBSEP
    for (j=1;;j++)
        if (a[i j])
            x = x verbatim(a[i j])
        else if (a[i j, 1])
            x = x "(" csexp(a, i j) ")"
        else
            break
    return x
}
function image(n,    url, caption) {
    url = imageurl(n)
    caption = imagecaption(n)
    if (!url) return ""
    return object("image", object("url", url) object("caption", caption))
}
function imageurl(n,    url, x) {
    if (!n) return ""
    if (name(n) == "img")
        url = verbatim(attr(n, "src"))
    # overwriting url selects the last one found
    x = imageurl(children(n))
    if(x) url = x
    x = imageurl(sister(n))
    if(x) url = x
    return url
}
```

```
function imagecaption(n,    caption) {
    if (!n) return ""
    if (name(n) == "figcaption")
         caption = hypertext(children(n))
    if (!caption) caption = imagecaption(children(n))
    if (!caption) caption = imagecaption(sister(n))
    return caption
}
function error(s) { print "error: " s; exit 1 }
```